\documentclass{ws-procs975x65}

\begin{document}

\title{QED Coupling  from a 5-D
Kaluza-Klein Theory}

\author{Andrea Marrocco}

\address{ICRA --- International Center for Relativistic Astrophysics}

\author{Giovanni Montani}

\address{ICRA --- International Center for Relativistic Astrophysics \\
Dip. Fisica, Universit\`a ``La Sapienza''
 Piazzale Aldo Moro 5, I-00185, Roma, Italy}

\maketitle

\abstracts{We discuss the possibility to obtain, 
from a five-dimensional free spinor Lagrangian,
the Quantum Electro-Dynamics (QED) coupling
via a Kaluza-Klein reduction of the theory.
This result is achieved taking a phase
dependence of the spinor field on the extra-coordinate and
modifying the corresponding connection.
The five-dimensional spinor theory is covariant under the
admissible coordinates transformations and its four-dimensional
reduction provides the QED coupling term.}

We assume, as a starting point,
a five-dimensional smooth 
manifold $V^4\otimes S^1$ 
endowed with a metric $\textbf{j}$ of 
components $j_{AB}$.\\
We make the hypothesis
that for all physical quantities there is no 
dependence on the fifth coordinate $x^5$; 
in particular for the metric we have that
$j_{AB}=j_{AB}(x^\mu)$, $(\mu=0,1,2,3)$.\\ 
Within a Kaluza-Klein theory the admissible
coordinates transformations reduce to the form
$x^{\mu ^{\prime }} = x^{\mu ^{\prime }} (x^{\nu }) \quad
x^{5^{\prime }} = x^5 + \alpha (x^{\nu })$,
according to which 
the component $j_{55}$ behaves like 
a scalar field and below it will be taken equal to unity.
Then, the components $j_{5\mu}$ can be
identified with a vector field
(the electromagnetic potential $A_\mu$) 
and the five-dimensional
Einstein-Hilbert action splits into the 
four-dimensional Einstein-Maxwell one.
In this theory, the components $j_{\mu\nu}$ differ 
from the four-dimensional 
metric tensor $g_{\mu\nu}$ by a term
$\propto A_{\mu }A_{\nu }$.

We investigate how to get,
via a splitting procedure, the
electrodynamics coupling term starting from a free spinor 
living on a Kaluza-Klein space-time. 
%
%
%; It is well known that the theory of a spinor field, on a 
%four dimensional curved space-time, requires to
%introduce a spinor connection $\Gamma _{\mu }$
%to make consistent
%the parallel transport for the Dirac's matrices. 
%So we define a new
%covariant derivative $D_\mu B$, where $B$ is a generic 
%field, which acts as           :
%
%\begin{equation}
   %D_\mu B\equiv \nabla_\mu B-[\Gamma_\mu,B]
   %\, ; 
   %\label{102}                                                                  %102%
%\end{equation} 
%
%it can be shown that $\Gamma _{\mu }$ has to take the form 
%
%\begin{equation}
   %\Gamma_\mu=-\frac{1}{4}
%   \gamma^\nu\nabla_\mu\gamma_\nu \label{103}                  %                    %103%
%\end{equation}
%
%and it follows the requested condition
%
%\begin{equation}
%   D_\mu\gamma_\nu=0
%   \, ; 
%   \label{104}                                             %104%
%\end{equation}
%
%both the above relations hold in any number of dimensions.\\
%
%For the spinor field
Let us introduce in
the five-dimensional space-time
a spinor field as a matter one, for which
we take the following Lagrangian density
(expressed in ``tetradic'' form) 
\begin{equation}
   ^5\Lambda=-\frac{i\hbar c}{2}\bar{\chi}
   \gamma^{(A)}D_{(A)}\chi+\frac{i\hbar c}{2}
   (D_{(A)}\bar{\chi})\gamma^{(A)}\chi+
   imc^2\bar{\chi}\chi \, ,
    \label{112}                                            %112%
\end{equation}
where $D$ is the new covariant derivative, 
endowed with the  spinor connection $\Gamma_\mu$
\begin{equation}
   \Gamma_\mu=-\frac{1}{4}
   \gamma^\nu\nabla_\mu\gamma_\nu \label{103}                                      %103%
\end{equation}
to preserve the parallel transport of 
Dirac's matrices.\\
Now we split the action associated to the 
four-dimensional quantities and achieve 
the dimensional reduction which provides the QED 
one. Firstly we express 
the spinor field $\chi$ and the `tetradic' spinor 
connection in terms of four dimensional quantities.
According to the cylindricity condition,
the spinor field is chosen as independent of
 the fifth coordinate, assuming the 
 standard form (\ref{103}) for the 
spinor connection i.e.
\begin{eqnarray}
\left\{\begin{array}{cc}
\chi=\frac{1}{\sqrt{L}}\psi
{}\\{}
{}\\{}
\bar{\chi}=\frac{1}{\sqrt{L}}\bar{\psi}
\label{1}
\end{array}\right.\end{eqnarray}                               %1%
where $\psi$ is the four-dimensional Dirac's
field, $L$ the extra-dimension length  and 
\begin{eqnarray}\left\{\begin{array}{cc}
\Gamma_{(\mu)}=^4\!\Gamma_{(\mu)}-\frac{ek}{4}
\gamma^{(5)}\gamma^{(\rho)}F_{(\mu)(\rho)}
{}\\{}
{}\\{}
\Gamma_{(5)}=-\frac{ek}{8}\gamma^{(\alpha)}
\gamma^{(\rho)}F_{(\alpha)(\rho)}\qquad\,
\label{111}
\end{array}\right.\end{eqnarray}                                                       %111%
being $F_{(\alpha)(\rho)}$ the electromagnetic
tensor, $\gamma^{(A)}$ the Dirac's matrices of 
flat space and  all the quantities have been expressed 
in the `tetradic' form.\\
Under these assumptions,  
we do not obtain the correct QED coupling; 
in fact the splitted action reads as 
\begin{align}
	S=\int\sqrt{-g}\left[-\frac{i\hbar }{2}
	\bar{\psi}\gamma^\mu D_\mu\psi+\frac{i\hbar }{2} 
	(D_\mu\bar{\psi})\gamma^\mu\psi+
	imc\bar{\psi}\psi+\frac{i\hbar ek}{8}
	\bar{\psi}\gamma^{(5)}\gamma^\mu\gamma^\nu\psi 
	F_{\mu\nu}\right]\,d^4x   \, ,
       \label{113}                       %113%
\end{align}
where $e$ is the electric charge and $k$ is a 
dimensional constant.
In (\ref{113}) there is no coupling between
the electromagnetic potential and the spinor current 
$J_\mu$, while it appears a new term that 
has no clear physical interpretation;
since such a term survives also on a flat space-time, 
then the present approach is wrong because it
predicts non-observed electrodynamics couplings.

Thus now we pursue an alternative point of view, 
observing that the spinor is a wave function and 
therefore admits a phase dependence on the fifth 
coordinate, such as 
\begin{equation}
  \chi\equiv\frac{1}{\sqrt{L}}
  e^{i\frac{2\pi x^5}{L}}\psi(x^\nu)\qquad
  \bar{\chi}\equiv
  \frac{1}{\sqrt{L}}e^{-i\frac{2\pi x^5}{L}}
  \psi(x^\nu). \label{88}                                             %88%
\end{equation}  
This choice for the spinor field is a good
one, as confirmed by considering that  
we obtain an equivalence between the charge operator 
and the fifth component of the five-momentum 
operator; it can be shown using the Noether's 
theorem. \\
It is just the phase term in the spinor field 
which allows the appearance of the expected  
QED coupling.\\ 
In spite of this success, we still have the problem
to avoid a misleading term; in fact the phase
introduces a contribution containing the
matrix $\gamma ^{(5)}$ in the splitted action.
This is eliminated
by choosing a different form for the spinor
connection as
\begin{eqnarray}
\left\{\begin{array}{cc}
\Gamma_{(\mu)}=^4\Gamma_{(\mu)}\equiv-
\frac{1}{4}\gamma^{(\rho)}
\gamma^{(\sigma)}R_{(\sigma)(\rho)(\mu)}
\qquad\qquad \mathrm{greek\,\,indexes
\,\,go\,\,from\,\,1\,\,to\,\,4}\qquad   
{}\\{}
\,\,\Gamma_{(5)}=\frac{2\pi i}{ L}\textbf{I}\qquad\qquad
\qquad\qquad\qquad\qquad\qquad 
\mathrm{with\,\,
\textbf{I}\,\,identity\,\,matrix}.
\, . 
\label{114}
\end{array}\right.\end{eqnarray}                                                %114%  
It is worth stressing that 
this form for the spinor
connection leaves the Lagrangian 
density (\ref{112}) invariant  under the restricted 
transformation of coordinates allowed in 
a Kaluza-Klein theory.\\
In terms of the new connection, 
the relations $D_A\gamma_B=0$
  no longer hold, 
but we preserve the validity of the four-dimensional conditions $ ^4D_\mu{}^4\!\gamma_\nu=0$.\\
By other words 
the Dirac's algebra is still valid in
the four-dimensional space-time once we have 
carried out the dimensional reduction on 
our action by integrating over the fifth dimension.\\ 
Working out the (\ref{112}) we obtain  
\begin{align}
	^5\Lambda=-\frac{i\hbar c}{2}
	\bar{\chi}\gamma^\mu{}^4\!D_\mu\chi-
	\frac{i\hbar c}{2}({}^4\!D_\mu\bar{\chi})
	\gamma^\mu\chi-\frac{2\pi ek\hbar c}{L}
	A_\mu\bar{\chi}\gamma^\mu\chi+
	imc^2\bar{\chi}\chi        \, . 
        \label{118}                             %118%
\end{align}
Starting from the five-dimensional action, integrating
on $x^5$ and using 
the cylindricity condition as well as the 
expression of $\chi$ and $\bar{\chi}$, we get
the total splitted action 
\begin{eqnarray}
	S=\frac{1}{c}\int\sqrt{-g}\bigg[
	-\frac{c^4}{16\pi G}\widehat{R}
	-\frac{1}{16\pi}F_{\mu\nu}F^{\mu\nu}-
\frac{i\hbar c}{2}\bar{\psi}
\gamma^\mu D_\mu\psi
{}\nonumber\\{}
+\frac{i\hbar c}{2}
(D_\mu\bar{\psi})\gamma^\mu\psi-
eA_\mu\bar{\psi}
\gamma^\mu\psi+
imc^2\bar{\psi}\psi\bigg]\,d^4x
\, , 
\label{122.a}                           %122.a
\end{eqnarray}
where we included also the part purely geometric  
of the action and taking 
\begin{eqnarray}\left\{\begin{array}{ll}
k=\frac{\sqrt{4G}}{ec^2}%\qquad\qquad\qquad
%\qquad\quad\,\,\,
%{}\\{}
{}\\{}
L=2\pi\sqrt{4G}\frac{\hbar}{ec}=
4.75\,10^{-31}\textrm{cm}
\, ; 
\label{124}                                              %124%
\end{array}\right.\end{eqnarray}
to get the terms
$\frac{1}{16\pi}F_{\mu\nu}F^{\mu\nu}$ 
and $eA_\mu\bar{\psi}\gamma^\mu\psi$ 
in the action.
The estimation of $L$ is
in agreement with other well-known ones
present in literature.   

%\section{Conclusions}

%In this paper we have shown a way to generate 
%the electrodynamics coupling 
%by the splitting of the five-dimensional action of a 
%free massive spinor field.\\
%We have seen how, providing the spinor field with 
%a phase dependence on the fifth coordinate, we are 
%able to produce the wanted term and how, assuming a 
%particular form of the spinor connection, we can 
%avoid the inconvenient of the appearence of a 
%term which hasn't any corresponding one in the 
%ordinary theory.\\ 
%We want to stress that the particular form of the spinor 
%field leads to the equivalence between the 
%translations along the fifth dimension and the gauge 
%transformations of the spinors associated to that of 
%the electromagnetic potential.\\
%The price to pay to obtain this results is the loss 
%of sense of the five-dimensional Einstein's 
%equations owing to the assumption of a non-standard 
%form of the spinor connection. Nevertheless we 
%don't know what is the theory in five dimensions.\\
%Our hope is to be able to generate by the splitting 
%of a multidimensional Kaluza-Klein theory's action 
%all the gauge couplings between the electro-weak 
%currents and the gauge fields. In order to do this we 
%have to provide with the right dependence on the 
%extradimensions all the leptonic fields. It isn't 
%easy as the case of only one extradimension because of 
%the non-commutativity of the transformation groups 
%acting on the fields.\\

\end{document}